%% file: main.tex
\documentclass[conference]{IEEEtran}
\IEEEoverridecommandlockouts
\usepackage{cite}
\usepackage{amsmath,amssymb,amsfonts}
\usepackage{mathtools, cuted}

\usepackage{algorithmic}
\usepackage{graphicx}
\usepackage{textcomp}
\usepackage{xcolor}
\usepackage{multirow}
\usepackage{caption}
\usepackage{subcaption}
\usepackage{comment}
\usepackage{url}
\usepackage{float}
\usepackage[T1]{fontenc}

\def\BibTeX{{\rm B\kern-.05em{\sc i\kern-.025em b}\kern-.08em
    T\kern-.1667em\lower.7ex\hbox{E}\kern-.125emX}}
\begin{document}

\title{Entanglement Distribution in Quantum Repeater with Purification and Optimized Buffer Time\\

\thanks{Author email addresses: yzang@uchicago.edu, xchen146@illinois.edu, atkolar@uchicago.edu, chungmiranda@anl.gov, sucharam@amazon.com, tzh@uchicago.edu, kettimut@anl.gov.
We thank Eric Chitambar for helpful discussion. This material is partly based upon work supported by the U.S. Department of Energy, Office of Science, National Quantum Information Science Research Centers. This work is also supported by Laboratory Directed Research and Development (LDRD) funding from Argonne National Laboratory, provided by the Director, Office of Science, of the U.S. Department of Energy under contract DE-AC02-06CH11357. This work is also supported by the National Science Foundation under award number DMR-2011854, and the NSF QLCI for Hybrid Quantum Architectures and Networks (NSF Grant No. 2016136).}
}

\author{
\IEEEauthorblockN{Allen Zang\IEEEauthorrefmark{1},
Xinan Chen\IEEEauthorrefmark{2},
Alexander Kolar\IEEEauthorrefmark{1},
Joaquin Chung\IEEEauthorrefmark{3},\\
Martin Suchara\IEEEauthorrefmark{4},
Tian Zhong\IEEEauthorrefmark{1},
Rajkumar Kettimuthu\IEEEauthorrefmark{3}}

\IEEEauthorblockA{\IEEEauthorrefmark{1}University of Chicago, Chicago, IL, USA,
\IEEEauthorrefmark{2}University of Illinois Urbana-Champaign, Urbana, IL, USA\\
\IEEEauthorrefmark{3}Argonne National Laboratory, Lemont, IL, USA,
\IEEEauthorrefmark{4}Amazon Web Services, Seattle, WA, USA\\
}}

\maketitle

\begin{abstract}
    Quantum repeater networks that allow long-distance entanglement distribution will be the backbone of distributed quantum information processing. In this paper we explore entanglement distribution using quantum repeaters with optimized buffer time, equipped with noisy quantum memories and performing imperfect entanglement purification and swapping. We observe that  increasing the number of memories on end nodes leads to a higher entanglement distribution rate per memory and higher probability of high-fidelity entanglement distribution, at least for the case with perfect operations. When imperfect operations are considered, however, we make the surprising observation that the per-memory entanglement rate decreases with increasing number of memories. Our results suggest that building quantum repeaters that perform well under realistic conditions requires careful modeling and design that takes into consideration the operations and resources that are finite and imperfect.
\end{abstract}

\input{introduction.tex}
\input{modeling.tex}

\input{results.tex}
\input{conclusion.tex}

\bibliographystyle{IEEEtran}
\bibliography{references}

\end{document}

%% file: introduction.tex
\section{Introduction}
The ability of quantum networks~\cite{kimble2008quantum,wehner2018quantum} to distribute entanglement will be necessary in order to perform distributed quantum information processing tasks such as distributed quantum computation~\cite{cuomo2020towards}, distributed quantum sensing~\cite{zhang2021distributed}, and quantum cryptography~\cite{pirandola2020advances}. 
Quantum repeaters~\cite{briegel1998quantum} are needed in order to build quantum networks because of lossy transmission of quantum states over long distances. 
However, errors that result in degradation of entanglement quality, coming from memory decoherence and noisy quantum gates and measurements, are inevitable in real-world implementations of quantum networks.
These errors must be considered in the evaluation of quantum network performance. 
The first-generation (1G) quantum repeater networks~\cite{muralidharan2016optimal} with probabilistic entanglement generation and purification will be the most suitable in the near term, for which we expect that each node in a quantum repeater network will contain a finite amount of noisy quantum memories and that physical operations including entanglement purification and swapping will be prone to errors. 
Therefore, the study of entanglement distribution with quantum repeaters under such conditions is of great practical importance.

Previous works have studied 1G quantum repeater protocol performance involving simplified optimization for subclasses of repeater protocols using specific physical operations~\cite{briegel1998quantum,dur1999quantum,childress2005fault,childress2006fault} or optimization using linear programming over a broader protocol space~\cite{jiang2007optimal}. 
Nevertheless, because of the complexity of quantum repeater network protocols (for discussion see~\cite{jiang2007optimal}),  incorporating all practical imperfections is difficult. The study of realistic 1G repeaters taking into account both imperfect quantum memory and entanglement purification has thus only started to emerge in recent years.
For instance, Goodenough et al.~\cite{goodenough2021optimizing} proposed a heuristic algorithm for optimization of repeater schemes; while both memory storage noise and entanglement purification were included, results were based on a simplified assumption of near-deterministic entanglement generation.
Brand et al.~\cite{brand2020efficient} addressed the problem of waiting time and fidelity distribution in quantum repeater chains by providing explicit algorithms to calculate those parameters; memory cut-off, an important tool for rate optimization, was not considered, however.
Li et al.~\cite{li2020efficient} included the cut-off time in secret-key rate optimization.
However, neither~\cite{brand2020efficient} nor~\cite{li2020efficient} included imperfect operations (entanglement swapping and purification), and the repeater protocols they considered were restricted to a tree-structured protocol stack.
 Laurenza et al.~\cite{laurenza2022rate} considered decoherence during both  storage and entanglement purification, but they focused on deriving an ultimate limit of network rate without worrying about specific implementation details.

Our contributions are threefold: 
\begin{enumerate}
    \item We extend the study of a quantum repeater chain architecture with hierarchically optimized (memory) buffer time~\cite{santra2019quantum} (see a brief introduction in~\ref{sec:architecture}), including multiple noisy quantum memories and potentially imperfect entanglement purification and swapping.
    \item We provide explicit analytical modeling of physical processes relevant to quantum repeater networks, namely, entanglement generation, purification, swapping, and memory decoherence-induced entanglement fidelity decay; these may be incorporated in future simulations.
    \item We demonstrate different and interesting phenomena of entanglement distribution performance under varying conditions, which both provide practical insight and can inspire future studies.
\end{enumerate}


The paper is organized as follows. In Sec.~\ref{sec:model} we provide explicit analytical modeling of relevant quantum operations and briefly review the architecture of a quantum repeater with buffer time.
In Sec.~\ref{sec:results} we explain the implementation of the numerical simulation and present our results.
We summarize our work in Sec.~\ref{sec:conclusion} and discuss future work.

%% file: modeling.tex
\section{System model}
\label{sec:model}

\subsection{Entanglement generation}
\label{sec:entanglement-generation}
Quantum memories on remote nodes have no direct physical interaction, which is typically required for entanglement generation between local subsystems through system evolution. 
Therefore, the generation of entanglement between memories on remote nodes is usually achieved through photon-mediated processes. 
In general, remote entanglement generation involves local photon-matter entanglement generation on both nodes to be entangled. This process is followed by photon interference to erase the ``which-way'' information at a middle station. This station will also communicate measurement results back to the two nodes, so that both ends know when entanglement generation is successful (called heralded entanglement generation).
A complete cycle of an entanglement generation attempt thus includes transmission of the photonic signal from two adjacent repeater nodes to the middle interference station and classical communication of measurement results from the middle station back to the repeaters. 
The ultimate lower bound of one entanglement generation cycle is $\tau=L_0/c$, where $L_0$ is the distance between neighboring nodes and $c$ is lightspeed. 
In practice, an entanglement generation cycle time will be constrained by the quantum memory (reuse) frequency, which can be on the order of $10~\mathrm{kHz}$~\cite{wu2021sequence}.

The initial fidelity of the entangled state upon successful generation (raw fidelity) is determined by all physical processes involved in the generation and is abstracted as a single parameter $F_0$. In this work we  focus on two specific families of noisy entangled states, namely, the two-qubit dephased Bell states $\rho_{dp}(F)=F\Phi^+ + (1-F)\Phi^-$ and the two-qubit Werner states (depolarized Bell states) \cite{werner1989quantum} $\rho_w(F)=F\Phi^+ + (1-F)(\Phi^-+\Psi^++\Psi^-)/3$, where $F$ is entanglement fidelity with respect to $\Phi^+$ state and $\Phi^+,\Phi^-,\Psi^+,\Psi^-$ are density matrices corresponding to four pure Bell states.
Similarly, the generation success probability $p_g$ will be generally determined by the properties of photon transmission channels, photonic coupling, and photon detectors. 
While the last two are local hardware and their imperfections can be described by a single combined hardware efficiency parameter $0\leq\eta_h\leq 1$, transmission performance is determined by channel length. 
Furthermore, since both remote nodes need to transmit their photonic modes, the success probability is proportional to the square of the transmission probability $\eta_t$. Thus, $p_g=\eta_t^2\eta_h$, where $\eta_t=\exp(-L/L_{att})$ is a function of channel length $L$ and the characteristic attenuation length for optical fiber is $L_{att}\approx 20~\mathrm{km}$.
We assume that the interference station is positioned in the middle of two repeater nodes, and thus the optical fiber between repeater node to the station is half the elementary link length $L=L_0/2$. We may therefore rewrite the entanglement generation success probability as $p_g=\exp(-L_0/L_{att})\eta_h$. 


\subsection{Entanglement swapping}
Entanglement swapping is performed on an intermediate repeater node, which holds two quantum memories that are respectively entangled with quantum memories on different remote nodes. 
A successful entanglement swapping will establish a longer entanglement ``link'' between the remote nodes. In reality, entanglement swapping will succeed with probability $p_s$ while imperfect physical operations will introduce additional errors to the extended entanglement link. 
We assume that entanglement swapping succeeds with probability $p_s$ and fails with probability $1-p_s$, in which case the two Bell pairs are assumed to be discarded. 
The assumed probabilistic nature requires classical communication of the success/failure of swapping to two end nodes, introducing another $\tau=L_0/c$ time.
When operations during entanglement swapping are perfect, Bell states will preserve their form after successful swapping; that is,   two dephased/depolarized Bell states will be swapped to a new dephased/depolarized Bell state. 
The fidelities of the new states after swapping are then
\begin{gather}
    F_{sw,dp} = F_1F_2 + (1-F_1)(1-F_2) \\
    F_{sw,w} = F_1F_2 + \frac{(1-F_1)(1-F_2)}{3}
\end{gather}


For imperfect operations, we consider two-qubit gate and single-qubit measurement error models as in~\cite{briegel1998quantum}, $\Tilde{U}_{ij}\rho\Tilde{U}^\dagger_{ij} = pU_{ij}\rho U^\dagger_{ij} + \frac{1-p}{4}I_{ij}\otimes\mathrm{tr}_{ij}\rho$ and $\Tilde{P}_{i=0,1} = \eta|i\rangle\langle i| + (1-\eta)|1-i\rangle\langle 1-i|$,
where $\mathrm{tr}_{ij}(\cdot)$ represents partial tracing over qubits $i,j$, $U_{ij}$ is an ideal qubit unitary, and $\Tilde{U}_{ij}$ is an imperfect implementation of $U_{ij}$, which has $p$ probability of perfect implementation and $(1-p)$ probability of resulting in  depolarizing error.
$\Tilde{P}_{i=0,1}$ is the POVM corresponding to imperfect implementation of single-qubit projective measurement $P_i=|i\rangle\langle i|$, which has $\eta$ probability of giving a correct measurement outcome. Note that although the POVMs alone cannot determine postmeasurement states, in our case the measured qubits are discarded (traced out), and we are only interested in the reduced state of the remaining subsystem. 
We provide an explicit formula of the fidelity of the swapped entangled state when the two input states are in Werner form,
\begin{equation}
\begin{aligned}
    & F_{sw,w} = \frac{1-p}{4} + p\left[\eta^2\left(F_1F_2 + 3e_1e_2\right)\right. \\
    & \left. +\,(1-\eta^2)\left(F_1e_2 + e_1F_2 + 2e_1e_2\right)\right],
\end{aligned}
\end{equation}
where we have defined $e_i=(1-F_i)/3$ to simplify notation. We note that the output state preserves Werner form under this specific operation error model.

\subsection{Entanglement purification}
Entanglement purification protocols 
take multiple copies of entangled states as input and output smaller amounts of entangled states upon success. 
Typically, a successful output will be of higher quality than the input states. 
In this work we consider specifically 2-to-1 recurrence protocols, namely, BBPSSW~\cite{bennett1996purification} and DEJMPS~\cite{deutsch1996quantum} with (almost) the same circuit. 
Moreover, as mentioned in Sec. \ref{sec:entanglement-generation}, we focus only on dephased Bell states $\rho_{dp}$ and Werner states $\rho_w$. 
We consider two imperfect Bell pairs with fidelities $F_1$ and $F_2$ as the input to the purification protocol. 
When no imperfect operations are considered, the output fidelity upon success and the success probability can be given in the following formulae: when input states are in dephased Bell state form,
\begin{align}
    F'_{dp}(F_1,F_2) &= \frac{F_1F_2}{F_1F_2 + (1-F_1)(1-F_2)}\\
    p_{dp}(F_1,F_2) &= F_1F_2 + (1-F_1)(1-F_2),
\end{align}
and when input states are in Werner form,
\begin{gather}
    F'_w(F_1,F_2) = \frac{F_1F_2 + \frac{(1-F_1)(1-F_2)}{9}}{F_1F_2 + \frac{F_1(1-F_2)+(1-F_1)F_2}{3} + \frac{5(1-F_1)(1-F_2)}{9}},
\end{gather}
\begin{align}
    p_w(F_1,F_2) &= F_1F_2 + \frac{F_1(1-F_2)+(1-F_1)F_2}{3}\\
    &+ \frac{5(1-F_1)(1-F_2)}{9} \nonumber
\end{align}


To demonstrate the effect of imperfections in the purification protocol, we use the same error models as for entanglement swapping. The major difference here is that for the considered purification protocol two CNOT gates are needed instead of only one for swapping. 
Here we provide explicit formulae of success probability and output fidelity upon success for purification of depolarized Bell states assuming identical gate success probability $p$ and measurement success probability $\eta$ on both repeater nodes:
\begin{align}
    p_{s,w} &= p^2[\eta^2+(1-\eta)^2](F_1F_2+F_1e_2+e_1F_2+5e_1e_2) \nonumber\\
    &+ 2p^2\eta(1-\eta)(2F_1e_2+2e_1F_2+4e_1e_2) + \frac{1-p^2}{2},
\end{align}
\begin{align}
    F_{s,w} &= 
    \frac{
    \begin{aligned}
        &[\eta^2+(1-\eta)^2](F_1F_2+e_1e_2)\\
        &+ 2\eta(1-\eta)(F_1e_2+e_1e_2) + \frac{1-p^2}{8p^2}
    \end{aligned}
    }{
    \begin{aligned}
        &[\eta^2+(1-\eta)^2](F_1F_2+F_1e_2+e_1F_2+5e_1e_2)\\
        &+ 2\eta(1-\eta)(2F_1e_2+2e_1F_2+4e_1e_2) + \frac{1-p^2}{2p^2}.
    \end{aligned}
    }
\end{align}
Note that different from swapping, the output state upon successful purification is not necessarily in Werner form, and twirling will be needed to transform the state back to Werner form.


\subsection{Fidelity dynamics under memory decoherence}
Entangled states are stored in quantum memories before being utilized later. Decoherence in quantum memories  results in the degradation of stored entangled states. As we study quantum repeater chain where end nodes do not have direct physical interaction, we consider an uncorrelated single-qubit error model,  $\rho_{AB}(t)=(\mathrm{E_A}\otimes\mathrm{E_B})[\rho_{AB}(0)]$, where $\mathrm{E_{A(B)}}[\cdot]$ is the quantum channel representing an error process on memory (qubit) A (B).
Specifically, for single-qubit dephasing (single Pauli) and depolarizing channels, we have the following analytical expressions of entanglement fidelity dynamics:
\begin{align}
    F_{dp}(t) 
    &= F_{dp}(0)\frac{1 + 2\beta^n}{3} + \frac{1 - \beta^n}{6},
    \label{eqn:fid_decay_dp}\\
    F_w(t) 
    &= F_w(0)\beta^n + \frac{1 - \beta^n}{4},
    \label{eqn:fid_decay_w}
\end{align}
where we have assumed that both quantum memories undergo a decoherence channel with the same amplitude characterized by decoherence rate $\kappa$, and we have further redefined the memory quality factor $\beta = \exp(-2\kappa\tau)$, while $n=t/\tau$ corresponds to time steps in simulation. Note that for the analytical formulae we have assumed that the initial state is also of dephased/depolarized Bell state form. 

\subsection{Nested repeater chain architecture with buffer time}
\label{sec:architecture}
Here we briefly describe the nested quantum repeater chain architecture with hierarchically optimized buffer time proposed in~\cite{santra2019quantum} and explain the inclusion of entanglement purification when multiple quantum memories are available. 
Consider a quantum repeater chain with $2^n+1$ nodes including two end nodes and $2^n$ elementary links of distance $L_0$. 
In a nested repeater architecture, Bell pairs are probabilistically generated over elementary links, and buffer time comes into play due to the probabilistic nature of entanglement generation; in a repeater with buffer time, entanglement generation can be re-attempted within every single (first level) buffer time and at the end of each buffer time if both sides of a first-level entanglement swapping station have received a Bell pair, entanglement swapping is performed. 
Conditioned on successful entanglement swapping, the entanglement link is extended to length $2L_0$ and is transferred to another set of quantum memories for the second nested level. 
In general, the successful entanglement swapping on level $i$ can be understood as a successful entanglement generation on level $i+1$, demonstrating that the nested architecture is self-similar (recursive). In this way, the buffer time for repeater level $n+1$ will in general be in unit of buffer time for repeater level $n$.
The process is hierarchically continued until the entanglement swapping on the highest nested level succeeds, which represents the successful distribution of one Bell pair between two end nodes of the repeater chain. 
The hierarchical buffer times can then be optimized by choosing buffer times that can maximize certain figure of merit, e.g. entanglement rate, on corresponding levels.
Furthermore, when multiple quantum memories are available for entanglement generation on one link, it is possible to generate multiple entanglement links during one buffer time, allowing entanglement purification. 
We assume that entanglement purification will be performed as soon as two entanglement links are generated, considering both success probability and potential fidelity improvement\footnote{Manuscript in preparation.}, and the later generated pair will be kept upon successful purification, which will offer higher fidelity when purification is imperfect.

Note that buffer time considered here is similar in spirit to cut-off considered in~\cite{li2020efficient} since both protocols are aimed at lowering the idling time of generated entanglement links. 
However, they are different in that the cut-off protocol examines the difference between generation times of any two entangled pairs whereas buffer time sets a fixed time range. 
We also emphasize that in the hierarchical buffer time architecture, entanglement purification is not strictly stacked upon entanglement generation before entanglement swapping since it is always possible that only one entanglement link is generated within one buffer time, which is different from the tree-structured repeater protocol considered in~\cite{brand2020efficient,li2020efficient} where the lower protocol must be successfully performed before the higher protocol starts. 

%% file: results.tex
\section{Results}
\label{sec:results}
We use an ad hoc numerical simulation\footnote{https://github.com/allenyxzang/Repeater-BufferTime} based on  the analytical modeling described in Sec.~\ref{sec:model} to study entanglement distribution in the simplest quantum repeater chain, namely, a first-level quantum repeater with two end nodes and only one middle swapping station node. 
It is noteworthy that for hierarchical repeater chain architectures even the first level analysis will offer valuable insight on the overall performance, as every level is self-similar (recursive).

\subsection{Simulation implementation}
\begin{figure}[htbp]
    \centering
    \includegraphics[width=\columnwidth]{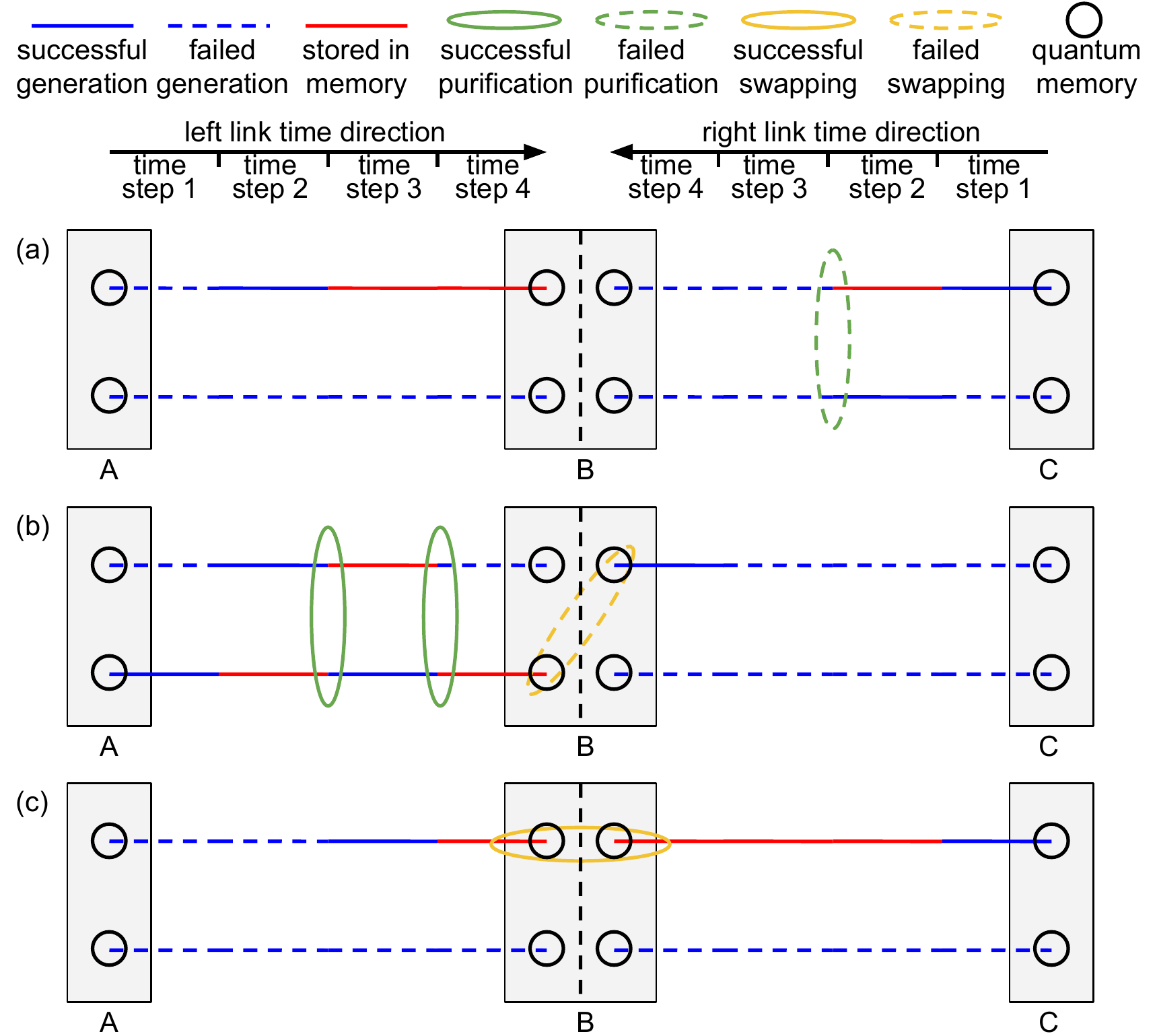}
    \caption{Illustration of different working scenarios for a repeater with buffer time. The first-level repeater involves two end nodes (A and C) and a middle swapping station B. Nodes A and C have two quantum memories, while node B has four, and we assume memory multiplexing for entanglement swapping. The buffer time is chosen to be equal to the time needed for 4 entanglement generation attempts. In (a) we show the probabilistic nature of entanglement purification, which gives rise to the possibility that after failed purification no more entanglement is generated before the end of the buffer time. In (b) the effect of probabilistic entanglement swapping is revealed, while demonstrating the possibility of entanglement pumping, that is, after successful entanglement purification the reinitialized memory can further generate new entanglement links to be purified. In (c) we show explicitly the difference between the repeater with buffer time and the repeater with tree-structured protocols; in other words,   entanglement swapping can be performed  whether or not any purification has been done.}
    \label{fig:schematics}
\end{figure}

We assume that each end node has $M$ quantum memories available for entanglement generation and that the middle station has $2M$ quantum memories. The $M=2$ scenario is illustrated in Fig.~\ref{fig:schematics}.

As a preliminary exploration of realistic scenarios with multiple available quantum memories, available entanglement purification, and imperfect quantum operations, in this work we fix a few system parameters: the length of one elementary link of repeater (distance between an end node and the middle node) is chosen to be optical fiber attenuation length, $L_0=L_{att}=20~\mathrm{km}$, which gives the photon loss factor $\eta_t=1/e\approx 0.37$ and one-way classical communication time $\tau_c=10^{-4}\mathrm{s}$.
For quantum memory we consider a modest reuse frequency of $1~\mathrm{kHz}$, which gives $\tau_m=10^{-3}~\mathrm{s}>10^{-4}~\mathrm{s}$, and thus we choose the entanglement cycle time as $\tau=10^{-3}~\mathrm{s}$.
We consider the quantum memory coherence time $\kappa^{-1}=1~\mathrm{s}$, which gives a memory quality factor $\beta = \exp(-0.002)\approx 0.998$. 
When imperfect operations are included, the operation success probability is assumed to be a modest value of 0.9. 
Regarding entanglement generation, in the following the raw fidelity of entangled state is assumed to be 1 corresponding to the best possible scenario.

The metrics of interest for repeater performance are the average entanglement distribution rate and fidelity distribution of successfully distributed states. The average entanglement distribution rate is $R_\mathrm{avg}$, which is defined as $R_\mathrm{avg} = p_\mathrm{succ}R(\Bar{\rho})/t_\mathrm{buffer}$,
where $p_\mathrm{succ}$ is the probability that an entangled state is successfully distributed between two end nodes within one buffer time, $R(\Bar{\rho})$ is the Rains (upper) bound~\cite{rains1999bound} of distillable entanglement for average distributed state with $\Bar{\rho}$ being the average of all successfully distributed states over $100,000$ repetitions of the simulation, 
 and $t_\mathrm{buffer}$ is the buffer time. 
Note that the Rains bound is tight for dephased Bell states and is dependent only on fidelity for Bell diagonal states: $R(\rho_\mathrm{BDS}(F)) = 1 - H_b(F)$, where $H_b(p) = -p\log p - (1-p)\log(1-p)$ is standard binary entropy function (according to Theorem 8 in~\cite{rains1999bound}). 
Since in our study the distributed states are in identical form, either dephased or depolarized Bell states, the fidelity of the average state is equal to the average of fidelity $\Bar{F}$.

For every specific choice of parameters, simulation of one buffer time will be repeated $100,000$ times to obtain statistics of average entanglement rate. 
The operation of repeater is discretized into time steps corresponding to the time required for each entanglement generation attempt. 
During one time step, the simulator will first check whether multiple entangled pairs are generated on the left and right links.
If so, entanglement purification will be performed on the generated pairs until at most one pair remains. 
Although entanglement purification also needs classical communication of results, we assume that the classical communication time is negligible, because purification is performed less frequently than entanglement generation. 
After this first phase, 
the rest of the available memories will attempt entanglement generation. 
The entanglement generation will potentially create new entangled pairs that enable entanglement purification, and similarly purification will be performed at the beginning of the next simulation time step. 
After $N_\mathrm{buffer}$ time steps the simulator will check the system state to evaluate the applicability of the final entanglement swapping. One final round of entanglement purification will be performed if needed to make sure that at most one pair remains on each side of the repeater. 
The probabilistic entanglement swapping will be performed as long as both sides have one entangled pair ready. If the swapping is successful, the fidelity of the distributed entangled pair will be recorded; otherwise an unphysical value of fidelity $-1$ will be recorded to indicate the failure of entanglement distribution in this trial. 
Note that the fidelity of a stored entangled pair matters only when it is involved in quantum operations. Therefore the entanglement fidelity will  be updated only before (to reflect the effect of memory decoherence) and after (to reflect the effect of operation) the application of entanglement purification and swapping.

\subsection{Advantage from additional physical resources}
\begin{figure}[htbp]
    \centering
    \begin{subfigure}[b]{0.51\linewidth}
        \centering
        \includegraphics[width=\linewidth]{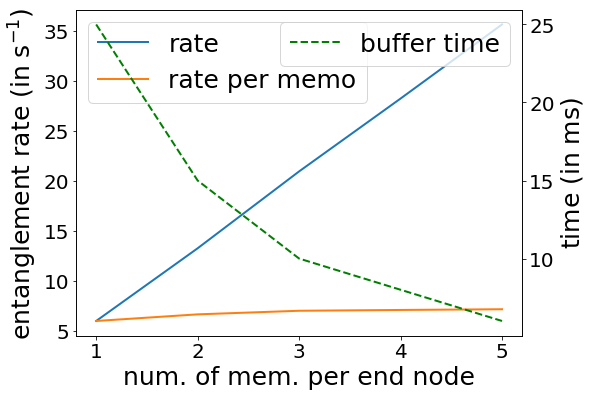}
        \caption{Optimal rate and buffer time}
    \end{subfigure}
    \hfill
    \begin{subfigure}[b]{0.47\linewidth}
        \centering
        \includegraphics[width=\linewidth]{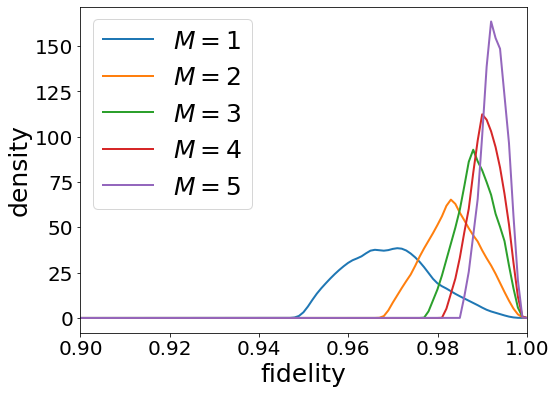}
        \caption{Fidelity probability density}
    \end{subfigure}
        \caption{Results for dephased Bell states ($\eta_h=0.1, p_s=0.5$)}
    \label{fig:dephased}
\end{figure}

\begin{figure}[htbp]
    \centering
    \begin{subfigure}[b]{0.51\linewidth}
        \centering
        \includegraphics[width=\linewidth]{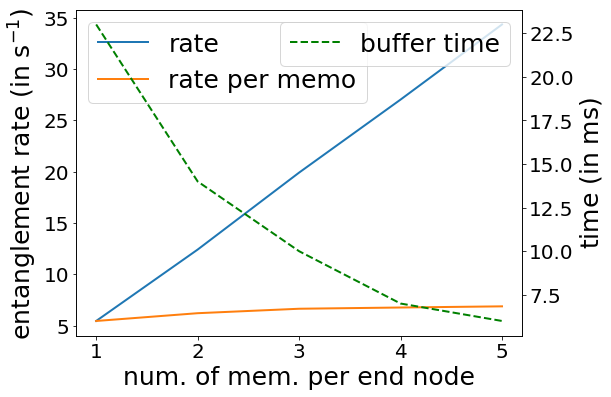}
        \caption{Optimal rate and buffer time}
    \end{subfigure}
    \hfill
    \begin{subfigure}[b]{0.47\linewidth}
        \centering
        \includegraphics[width=\linewidth]{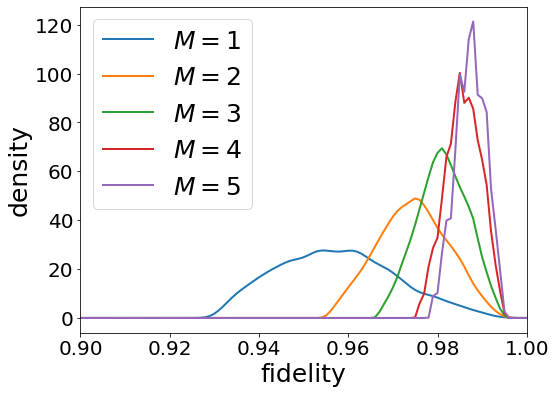}
        \caption{Fidelity probability density}
    \end{subfigure}
        \caption{Results for Werner states ($\eta_h=0.1, p_s=0.5$)}
    \label{fig:werner}
\end{figure}

In this section we present our simulation results on the optimal entanglement distribution rate and optimized buffer time versus the number of available quantum memories. Additionally, we also present the probability density of entanglement fidelity conditioned on successful distribution. 
Here no imperfect operations are considered, and memory size on each end node varies from 1 to 5, with entanglement generation hardware efficiency $\eta_h=0.1$ and swapping success probability $p_s=0.5$. 
We note here that although the process of buffer time optimization is not explicitly demonstrated, it is achieved by evaluating entanglement distribution rate for buffer time (in unit of entanglement generation cycle time) $N_\mathrm{buffer}\in\mathbb{Z}^+$ ranging from 1 to 30 (30 here is only a cutoff for numerics and is justified by presented results as optimal buffer times are all below 30) and then choosing the buffer time which maximizes entanglement distribution rate.
The results for dephased Bell states are shown in Fig.~\ref{fig:dephased} while those for Werner states are presented in Fig.~\ref{fig:werner}. 

Qualitatively, we can see that for the two forms of noisy Bell states, both the  optimal entanglement rate and optimized buffer time vary as the number of quantum memories varies, whereas for the fidelity probability density they  are similar.
We can also observe that the optimal rates for dephased Bell states are higher than for Werner states and that the peaks of fidelity density for dephased Bell states are also closer to one. 
These phenomena demonstrate that entanglement decays faster when the decoherence channel is depolarizing rather than dephasing when the decoherence amplitude $\kappa$ is fixed, as seen from Eq.~\ref{eqn:fid_decay_dp} and~\ref{eqn:fid_decay_w}. 
Furthermore, the figures of merit demonstrate interesting behaviors with varying number of available quantum memories. 
The decaying optimized buffer time as available memories increase reveals the effect of multiplexing:  when more memories can attempt entanglement generation, less time is needed for successful generation. 

As expected, when more memories are available entanglement purification is possible, and the entanglement rate increases. More interesting is the increase of entanglement purification rate per quantum memory. 
The latter phenomenon suggests that combining additional physical resources (memories and thus entanglement purification) results in additional advantage in comparison with simply in-parallel operation of single-memory no-purification repeaters with the same amount of memories. 
Additionally, since realistic distributed quantum information processing tasks will always set certain requirements, especially lower threshold, on the entanglement fidelity of each distributed entangled pair to be utilized, the probability density function of entanglement fidelity upon successful distribution demonstrates the beneficial effect of incorporating more physical resources, in that the peak of probability density function becomes closer to one, suggesting higher probability of distributing high-fidelity entanglement.

\subsection{Effect of imperfect operation}
\begin{figure}[htbp]
    \centering
    \begin{subfigure}[b]{0.52\linewidth}
        \centering
        \includegraphics[width=\linewidth]{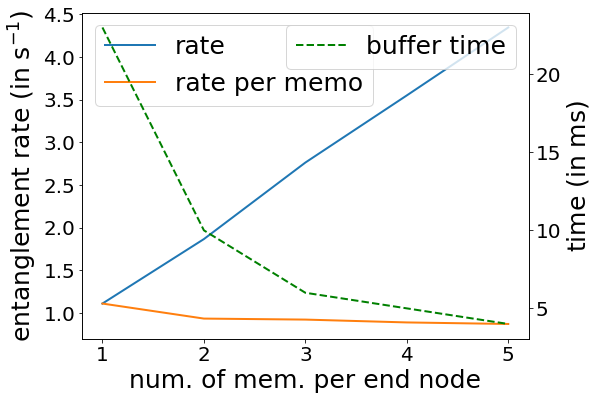}
        \caption{Optimal rate and buffer time}
    \end{subfigure}
    \hfill
    \begin{subfigure}[b]{0.46\linewidth}
        \centering
        \includegraphics[width=\linewidth]{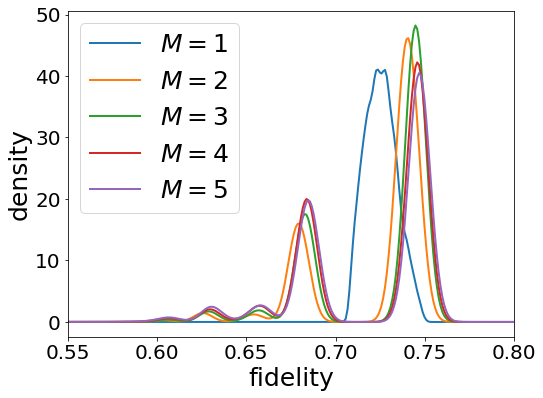}
        \caption{Fidelity probability density}
    \end{subfigure}
        \caption{Results for Werner states with operation imperfection ($\eta_h=0.1, p_s=0.5,p_\mathrm{gate}=p_\mathrm{meas}=0.9$)}
    \label{fig:werner_imperfect}
\end{figure}

In comparison with the preceding section where no imperfect operations are included, here we consider two-qubit gate and single-qubit measurement success probability $p_\mathrm{gate}=p_\mathrm{meas}=0.9$, without changing other parameters for the Werner state.
Fig.~\ref{fig:werner_imperfect} shows the simulation results.

We  observe that the trends of optimal entanglement rate and optimized buffer time with varying memory number do not change qualitatively;  the  optimal entanglement rate still increases when more memories are available, while the optimized buffer time decreases. However, an important difference is in the behavior of the entanglement rate per memory. 
The per-memory rate surprisingly falls below the single-memory rate when more memories are incorporated, which means that now simply operating the single-memory repeater in parallel will outperform combining available memories together and including imperfect entanglement purification. 
Additionally, we can obtain further information about the imperfect repeater from the fidelity probability density. 
We observe more than one peak for multiple memories, which represent different times of entanglement purification that were performed. 
Notably, the highest peak for multiple memories is still closer to one when compared with the one-memory case, which corresponds to the case when no purification is performed. 
When purification is performed, however, such cases correspond to lower peaks that are farther from one, demonstrating the harmful effect of imperfect operations to quantum repeater.

%% file: conclusion.tex
\section{Conclusion}
\label{sec:conclusion}
In this paper we study entanglement distribution in a simple quantum repeater chain with optimized buffer time that has only three nodes with a few noisy quantum memories and allows entanglement purification. 
We demonstrate the additional advantage in per-memory entanglement distribution rate that can be obtained by incorporating more quantum memories and including entanglement purification when all operations are perfect even under memory decoherence, compared with parallel operation of separate repeaters. 
 Surprisingly, however, we observe that this advantage is lost when imperfect operations are considered. 
Our results demonstrate the existence of rich and interesting phenomena in quantum repeaters with finite, imperfect resources, even in the simplest setup, and offer insight on near-term, real-world implementation of quantum repeater networks. 

The results reported in this paper are preliminary, and we will perform further detailed analysis on broader parametric spaces to explore and demonstrate richer phenomena and will provide more comprehensive explanations  based on additional theoretical models.